\documentclass[a4paper,11pt]{article}
\usepackage{pos}

\title{A physicist-friendly reformulation of the mod-two Atiyah-Patodi-Singer index}
\author[a]{Hidenori Fukaya}
\author[b]{Mikio Furuta}
\author*[a]{Yoshiyuki Matsuki}
\author[c]{Shinichiroh Matsuo}
\author[a]{Tetsuya Onogi}
\author[a]{Satoshi Yamaguchi}
\author[d]{Mayuko Yamashita}

\affiliation[a]{Department of Physics, Osaka University, Osaka, Japan}

\affiliation[b]{Graduate School of Mathematical Sciences, The University of Tokyo, Tokyo, Japan}

\affiliation[c]{Graduate School of Mathematics, Nagoya University, Nagoya, Japan}

\affiliation[d]{Research Institute for Mathematical Sciences, Kyoto University, Kyoto, Japan}

\emailAdd{hfukaya@het.phys.sci.osaka-u.ac.jp}
\emailAdd{furuta@ms.u-tokyo.ac.jp}
\emailAdd{ymatsuki@het.phys.sci.osaka-u.ac.jp}
\emailAdd{shinichiroh@math.nagoya-u.ac.jp}
\emailAdd{onogi@phys.sci.osaka-u.ac.jp}
\emailAdd{yamaguch@het.phys.sci.osaka-u.ac.jp}
\emailAdd{mayuko@kurims.kyoto-u.ac.jp}

\abstract{Gauge anomaly in 4-dimensions can be viewed as a current inflow into an extra-dimension, where the total phase of the fermion partition function is given in a gauge invariant way by the Atiyah-Patodi-Singer(APS) eta-invariant of a 5-dimensional Dirac operator. However, this formalism requires a non-local boundary condition, with which the physical roles of edge/bulk modes are unclear and how the causality of the theory is maintained is not obvious. In this work, we consider a special case where the Dirac operator is in a real representation and its eta invariant becomes the mod-two type APS index. We propose a physicist-friendly reformulation of the mod-two index using domain-wall fermion formalism, which naturally describes how the global anomaly is canceled between edge and bulk.}

\FullConference{%
 The 38th International Symposium on Lattice Field Theory, LATTICE2021
  26th-30th July, 2021
  Zoom/Gather@Massachusetts Institute of Technology
}

\begin{document}
\maketitle

\section{Introduction}
%%%%アノマリーの重要性
%量子アノマリーはQFTにおける対称性や保存則を十分に理解する上でとても重要な概念であり、これはていえねるぎーの物理を予言する。
%例えば、グローバル対称性 がアノマラスである時、tHooft anomaly matchingでどうのこうの。
Quantum anomaly \cite{Adler:1969gk,Bell:1969ts} have played important roles in particle physics. The anomaly of the axial U(1) symmetry describes some low-energy behavior of QCD, like $\pi^0\to \gamma\gamma$ decay. The gauge anomaly restricts particle contents of the theory beyond the standard model. The anomaly matching \cite{tHooft:1979rat} is a powerful tool for the analysis of strongly coupled theory even when the corresponding gauge symmetry does not exist in the theory.
% is a crucial concept in fully understanding low-energy physics. For instance, the description of the decay rate for $\pi^0 \to \gamma \gamma$ is a typical example of it. Also, the idea of 't Hooft anomaly matching \cite{tHooft:1979rat} gives the limitation for the energy spectrum in the infra-red region in QCD. 
 %%%%%アノマリーインフローが流行っている
%いっぽう、ゲージ対称性がアノマラスであった場合は問題である。なぜなら、ゲージ理論においてユニタリーせいや繰り込み可能性を司っているから。
%しかし、この場合でもhigher-dimensionにアノマラスな理論を埋め込むことによって、ゲージ対称性を回復できる可能性がある。For instance、4次元を5次元に埋め込んでどうのこうの。これはアノマリーインフローメカニズムと呼ばれている。この考え方は素粒子論ではこういうところに使われていて、物性でもこういうところで使われている。摂動論を強調。Quantum anomaly [1,2] have played important roles in particle physics.

Any gauge anomaly must be cancelled for the theory to be consistent. The cancellation of anomaly among quarks and leptons in the standard model is the most well-known example. In this work, we consider another type of anomaly cancellation by embedding the anomalous fermion into a higher-dimensional theory. For example, let us consider a four-dimensional chiral gauge theory, which has a perturbative gauge anomaly. The anomaly can be viewed as a current flowing into the fifth dimension from the four-dimensional edge, which is cancelled by the Chern-Simons action in the bulk \cite{Callan:1984sa}. This anomaly inflow or matching can be viewed as a realization of the bulk-edge correspondence of the fermion system in a topological phase \cite{BEChatsugai}.
%When the gauge symmetry is anomalous, it is seriously critical. It's because the gauge symmetry guarantees the unitarity and the renormalizability in the gauge theory. However, it is possible to recover the gauge symmetry by embedding the anomalous theory into the higher-dimensional theory. For example, let us consider a four-dimensional chiral gauge theory, which has a perturbative gauge anomaly. The anomaly can be viewed as a current flowing into a five dimension, but conserved in total five-dimensional Chern-Simons theory. This is well-known as the anomaly inflow mechanism \cite{Callan:1984sa}. It exactly agrees with the bulk-edge correspondence in condensed matter physics \cite{BEChatsugai}. 

%%%%%%グローバルアノマリーでもアノマリーインフローが指摘
In \cite{Witten:2015aba}, it was pointed out that the anomaly inflow mechanism can also be generalized to the global anomaly \cite{Witten:1982fp} along with perturbative anomalies, that is, a global anomaly in a $d$-dimensional theory, where it is absorbed into a $d+1$-dimensional bulk \cite{Witten:1985xe,Witten:2016cio}. Concretely, the anomaly inflow is extended by an $\eta$-invariant of Atiyah, Patodi, and Singer (APS) \cite{APS}. The relation between the $\eta$-invariant and anomalies was mathematically formulated by Dai and Freed \cite{Dai:1994kq} (the physical formulation was given in \cite{Yonekura:2016wuc}).
%最近、このアノマリー印フローがグローバルアノマリーをふくむあのまりーに他強いても有効であることが指摘された。
%すなわち、あるd次元のグローバルアノマリーは、d+1次元の多様体をくっつけたものを考えると、トータルの分配関数はゲージ不変であるというものである。

%一般にぶんぱいかんすうの位相はエータ不変量と呼ばれる量で表される。

%%%%%%%これらの文献ではAPS境界条件が使われている。
%これらの文献において、このエータ不変量はAPS境界条件というが使用されている。これはnon-localな境界条件であり、エッジモードの存在が許されない。そのため、ばるくとエッジの寄与がぶんりされておらず、バルクーエッジ対応のような関係は明らかではない。
%In the articles \cite{Dai:1994kq,Freed:2014iua}, the $\eta$-invariant is defined with APS boundary conditions. However, this boundary condition troubles physicists because it is non-local. Moreover, it does not allow for the existence of edge modes, thus, it is quite difficult to separate the $\eta$-invariant into bulk and edge contributions as in the usual bulk-edge correspondence.

The eta invariant introduced in Refs. \cite{Dai:1994kq,Freed:2014iua} is, however, not very physicist-friendly. It imposes the so-called APS boundary condition, which is non-local\footnote{In \cite{Witten:2019bou}, it was discussed how any unphysical properties caused by the APS boundary condition are cancelled in the fermion partition function.}. Moreover, the boundary condition does not allow the edge-localized modes to exist, and therefore, it is difficult to separate the bulk and edge contributions.

The alternative set-up without any nonlocal condition was first proposed in \cite{Fukaya:2017tsq}
by three physicists of the authors. Instead of the manifold with the boundary, they considered a domain-wall fermion Dirac operator on a flat 4-dimensional manifold extended from the original boundary, where the sign of the mass is set to flip. They perturbatively showed that the $\eta$ invariant of the domain-wall Dirac operator of a pseudoreal fermion coincides with the APS index\footnote{The APS index of a pseudoreal Dirac operator in even dimensions coincides with the $\eta$ invariant mod integer.}.
In the new formulation, the contributions from the bulk and edge modes are manifest, and the anomaly of the time-reversal symmetry is cancelled between them. Moreover, a local boundary condition is automatically imposed. In \cite{Fukaya:2019qlf}, three mathematicians of the authors joined and gave a
mathematical proof that this reformulation of the APS index is valid on any even-dimensional curved manifold. In \cite{Fukaya:2019myi}, the work was extended to a definition of the APS index in lattice gauge theory. For a recent review, we refer the readers to \cite{Fukaya:2021sea}.

%For real fermions, the $\eta$-invariant is known as mod-two APS index \cite{Atiyah:1971rm}. The problem for real fermions is even more difficult than that for pseudo-real fermions. In the case of pseudo-real fermions, the APS index theorem \cite{APS} tells us the bulk and edge contributions explicitly. However, for real fermions, the natural separation into the bulk and edge contributions has not been known. Furthermore, the mod-two APS index is closely related to the Witten's global anomaly \cite{Witten:1982fp} and requires a non-perturbative treatment.

%In this work, we reformulate the mod-two APS index in the physicist-friendly setup with the domain-wall fermion. Furthermore, we provide the perspective as the bulk-edge correspondence for the Witten's $SU(2)$ global anomaly. This proceedings is based on our recent work \cite{Fukaya:2020tjk}.

In this work \cite{Fukaya:2020tjk}, we consider an extension to real fermions on an odd-dimensional manifold. When the Dirac operator is real, its $\eta$ invariant coincides with the mod-two APS index. The mod-two APS index is more difficult compared to the standard APS, in that a natural separation of the bulk and edge modes is not obvious in the index formula, and it essentially requires a non-perturbative treatment to describe the global anomaly. We will show below that domain-wall fermion can describe the mod-two APS index and even offers a natural separation of the bulk and edge contributions.

\section{Short review of Global anomaly and Mod-two APS index}
In this section, we briefly review the $SU(2)$ global anomaly and the mod-two AS index following \cite{Witten:1982fp}. Then, we discuss its extension to the mod-two APS index on a general manifold with boundary.

\subsection{Global anomaly and Mod-two Atiyah-Singer(AS) index}
In \cite{Witten:1982fp}, it was pointed out that the gauge theory in the fundamental representation of the gauge group $SU(2)$ with an odd number of Weyl fermions is inconsistent. 
In general, when the Dirac operator is real and anti-symmetric, the same inconsistency exists. %一般的には実かつ反対称なDirac演算子に対しても同様のinconsistentが存在する。

Let us consider a real and anti-symmetric Dirac operator $D_{X}$ on a manifold $X$ and assume that there are no zero eigenvalues. The non-zero eigenvalues of $D_{X}$ make a complex conjugate pair $\pm i \lambda$, where $\lambda$ is real. In this setup, it is shown that the Weyl partition function is real. Therefore, there is no perturbative anomaly since the anomaly appears in the complex phase in the partition function.   

To see the non-perturbative anomaly, we consider the smooth connection between the two gauge equivalent configurations $A_{\mu}$ and $A_{\mu}^g$, i.e., $A^s_{\mu} = (1-s)A_{\mu} + sA^{g}_{\mu},\ (0\leq s \leq 1)$. Here $A^{g}$ is a gauge field which is performed the gauge transformation in the nontrivial class of $\pi_{4}(SU(2))$. In this case, the eigenvalues flip the sign by odd times and the spectral flow against $s$ is odd. Therefore, the sign of Weyl partition function is not determined in a gauge invariant way. 

The proof is given by the mod-two Atiyah-Singer(AS) index on a five-dimensional manifold. Let us consider a five-dimensional cylinder $S^{4}\times R$, where $R$ is the fifth direction $s$. Since the both ends of the cylinder are gauge equivalent, we can construct the torus $S^{4}\times S^{1}$, which is called a mapping torus. On the mapping torus, the number of zero modes of the real Dirac operator $D$ mod $2$ is known as the mod-two AS index $I_{AS}^{mod-two}(D)$. In \cite{Witten:1982fp}, it was proved that the spectral flow of original four-dimensional $D_X$ is odd if the mod-two AS index is odd. Moreover, it was also shown that the mod-two AS index is odd when the gauge transformation $g$ is in the nontrivial class of $\pi_{4}(SU(2))$.

\subsection{Non-perturbative anomaly inflow and Mod-two APS index}
%%%%%%mod-two APSの話。
It is well-known that perturbative anomalies in $d$-dimensional spacetime is related to the Chern-Simons action in $(d+1)$-dimensional spacetime \cite{Zumino:1983ew,Jackiw:1983nv,Stora:1983ct}. It was formulated as the anomaly inflow mechanism: the anomaly on the boundary is viewed as the current flowing into the bulk \cite{Callan:1984sa}. 
%アノマリー流入が古くから知られている。さらに、non-perturbativeでもできることが示唆され、正当化された。具体敵には、d+1次元境界付きmfdの分配関数の位相はAPS境界条件という境界条件んの元でeta-invariantという量で表され、以下で定義される。
%これはゲージ不変であり、anomaly-inflowの見方を与えている。これらのformulationはdai-Freed定理に基づいている。
%%%%
%Dirac operatorが実の時、eta0invariantはmod-two APSと呼ばれ、上の定義より、zeromodoのみの輪である。

Recently, the anomaly inflow was generalized and extended to global anomalies\cite{Witten:2015aba,Dai:1994kq,Witten:2019bou}. The anomaly of a $d$-dimensional fermion can be absorbed into a $(d+1)$-dimensional bulk and  the phase of the total fermion determinant is described by exp$(i\pi \eta(iD))$, where the $\eta$-invariant of the Dirac operator $D$ on the $(d+1)$-dimensional manifold is defined by
\begin{eqnarray}
\eta(iD)=\sum_{n}\text{sgn}\lambda_{n}+h,
\end{eqnarray}
where the summation is regularized, $\lambda_{n}$ is the non-zero eigenvalues of  $iD$, and $h$ is the number of the zero-modes. This quantity is gauge-invariant and includes non-perturbative effects as well as the Chern-Simons action. 

When the Dirac operator is real, the $\eta$-invariant equals to the mod-two APS index \cite{Atiyah:1971rm} which is given by the number of the zero-modes $h$ since the non-zero eigenvalues make a pair. In this case, It implies the anomaly inflow for the global anomaly. 

We note that the mod-two APS indices are defined on any $(d+1)$-dimensional manifolds with boundary, no longer restricted to a mapping torus. In \cite{APS,Dai:1994kq,Freed:2014iua}, the so-called APS boundary condition was imposed to make the quantity $\eta(iD)$ well-defined  keeping the Dirac operator massless. However this boundary condition is non-local and physicist-unfriendly in that no edge-localized modes survive, which is quite different from our target physical setup of topological insulators.

%Witten and Yonekura described that the APS boundary conditions are harmless to be used in physical setups \cite{Witten:2019bou}. 
%In their setting, The coordinate of the extra-direction is considered as time, and the APS boundary condition is introduced as a "state" in performing path-integration. They showed that the unphysical nature in the partition function is canceled out.

%However, the question remains as to why the APS boundary condition should be used when considering physical situations in the first place. And as long as the APS boundary condition is used, the contributions of bulk and edge are not clear. 

%To this question, we have shown that the domain-wall fermion is helpful in the case of pseudo-real fermions \cite{Fukaya:2017tsq,Fukaya:2019qlf}. The reformulation with the domain-wall fermions provides a local boundary condition and ****clarifies the bulk and edge contributions. 

%For real fermions, the $\eta$-invariant is known as mod-two APS index \cite{Atiyah:1971rm}. The problem for real fermions is even more difficult than that for pseudo-real fermions. In the case of pseudo-real fermions, the APS index theorem \cite{APS} tells us the bulk and edge contributions explicitly. However, for real fermions, the natural separation into the bulk and edge contributions has not been known. Furthermore, the mod-two APS index is closely related to the Witten's global anomaly \cite{Witten:1982fp} and requires a non-perturbative treatment.

In this work \cite{Fukaya:2020tjk}, we give a reformulation of  the mod-two APS index
using the domain-wall fermion. Our formulation is physicist-friendly and requires no unphysical boundary condition.
Moreover, it offers a natural separation of the bulk and edge contributions of global anomaly so that one can easily see how they are canceled, which is a realization of the bulk-edge correspondence.

\section{Domain-wall fermion and Mod-two APS index}
In this section, we describe the reformulation of the mod-two APS index with the domain-wall fermion. The mathematical rigorous discussion is given in \cite{Fukaya:2020tjk}.
%いろいろ定義を述べる。

First, let us explain some definitions and setting. As we mentioned in the previous section, the mod-two index is defined by the number of zero-modes of a real anti-symmetric Dirac operator $D$,
\begin{eqnarray}
\text{Ind}(D):= \text{dim\,Ker}D\ \ (\text{mod} 2).
\end{eqnarray}

Let $Y$ be a closed Riemannian manifold and $X \subset Y$ be a submanifold that separates $Y$ into two compact manifolds $Y_{+}$ and $Y_{-}$ with the common boundary $X$. We assume that near the boundary $X$, the manifold has the collar structure $[0,+\epsilon)_{s}\times X$, and the Dirac operator $D$ is assumed to take the form 
\begin{eqnarray}
D=\gamma^{s}\partial_{s}+D_{X},
\end{eqnarray}
where $s$ is the coordinate normal to the boundary, and $\gamma^{s}$ is the gamma matrix which plays a role of a chiral operator. The gamma matrix $\gamma^{s}$ satisfies the relations $(\gamma^{s})^2=1$ and $\{\gamma^{s},D_{X}\}=0$.

The APS boundary condition eliminates the positive eigenstates on the boundary $X$,
\begin{eqnarray}
(A+|A|)\phi|_{X}=0,
\end{eqnarray}
where $A=\gamma^{s}D_X$ and $|A|$ is the absolute value of the operator $A$. This boundary condition is non-local since the operator $|A|$ is a non-local operator.

%本研究の目的は、APS境界条件のもとでのmod-two APSindexとドメイヌォールフェルミオン分配関数の位相 (mod2)が等しいこと、すなわち、いかをしめすことである。
The purpose of this study is to show that the mod-two APS index with APS boundary condition and the sign of the partition function of the domain wall fermion are equivalent:
\begin{eqnarray}
\label{eq:modtwoDW}
\text{Ind}_{\text{APS}}(D|_{Y_{+}})= \frac{1-\text{sgn\,Det}(D_{\text{DW}}D_{\text{PV}}^{-1})}{2}\ \ (\text{mod} 2),
\end{eqnarray}
where $D_{DW}$ and $D_{PV}$ on the closed manifold $Y$ are defined as follows. 
\begin{eqnarray}
&&D_{DW}=D+M\epsilon(s),\\
&&D_{PV}=D-M.
\end{eqnarray}
The function $\epsilon(s)$ is the sign function, i.e., $\epsilon(s)=+1$ for $s>0$ and  $\epsilon(s)=-1$ for $s<0$. 
\subsection{Some theorems used in our claim}

In this subsection, we introduce two theorems that are necessary to prove our main theorem. We will leave their proofs to these references \cite{APS,Carey,Fukaya:2019qlf,Fukaya:2020tjk}.

We consider the cylindrical manifold $Y_{\text{cyl}}$ which is $Y_{+}$ with semi-infinite cylinder attached to the boundary $X$, i.e., $Y_{\text{cyl}}:=Y_{+}\cup(-\infty,0]\times X$. In this setting, the mod-two APS index on $Y_{+}$ is equivalent to the mod-two AS index on $Y_{\text{cyl}}$ \cite{APS}.
\begin{eqnarray}
\label{eq:indexcylinder}
\text{Ind}_{\text{APS}}(D|_{Y_{+}})=\text{Ind}(D_{\text{cyl}}).
\end{eqnarray}

%位置依存する質量高を持つd+1次元のDirac 演算子のインデックスはd次元のindexとmassless modeによる1次元 indexの席で与えられる。

%6次元座標tに依存する実反対称なDirac 演算しにおいて、
%(t=-1-1)まで変化させたときに、jペクトラルフローすなわち、固有値のペアが何回横切るかは以下の公式で与えられる。

Let us consider a one-parameter family of the real anti-symmetric
Dirac operator $D_t$ labelled by t. When t moves from $-1$ to $1$, the spectral flow of $D_t$, i.e., how many times pairs of eigenvalues cross zero, is given by the following formula \cite{Carey}
\begin{eqnarray}
\label{eq:spectral}
\text{Sf}(D_{-1},D_{1})=\frac{1-\text{sgn\,Det A}}{2},
\end{eqnarray}
where the operator $A$ satisfies the relation $D_{1}=A^{\dagger}D_{-1}A$.

%この節では我々のメインの主張である、domain-wall fermionを用いたmod-two APSの再構築について紹介する。初めにdomain-wall femrionを用いたreformulationとAPSboundary conditionで定式化したものが等しいことを示す。そして、mod-two APS指数定理のアノマリー流入としての見方を紹介する。これによってグローバルありーのバルクエッジ対応の見方が明らかになる。

\subsection{Proof of our main theorem}
Now, we give the proof of our main theorem (\ref{eq:modtwoDW}). 

We consider a $6$-dimensional Dirac operator
\begin{eqnarray}
\label{eq:dirac6}
\widehat{D}_{t}:=\left(\begin{array}{cc}
\partial_{t}&D+M\kappa(t)\\
D-M\kappa(t) &-\partial_{t}
\end{array}
\right),
\end{eqnarray}
where $t$ is the $6$th direction, $\kappa(t)=\epsilon(s)$ for $t>0$ and $\kappa(t)=-1$ for $t<0$ (See Fig.1). $\widehat{D}_{t}$ can be decomposed into the $5$-dimensional Dirac operator and $1$-dimensional Dirac operator with domain-wall mass,
 \begin{eqnarray}
\widehat{D}_{t}&=&\left(\begin{array}{cc}
0&D\\
D&0
\end{array}
\right)+D^{1D}\nonumber\\
D^{1D}&:=&\left(\begin{array}{cc}
\partial_{t}&M\kappa(t)\\
-M\kappa(t) &-\partial_{t}
\end{array}
\right)
\end{eqnarray}
 
Note that the mod-two index of $\widehat{D}_{t}$ is expressed as the product of the index of $D$ and that of $D^{1D}$. In fact, we can consider the zero-modes $\psi$ of $D$, the zero-mode $u\text{exp}(-M|t|)$ of $D^{1D}$ independently, where $u=(1,1)^{T}$. And the zero modes of $\widehat{D}_{t}$ are expressed as these product $u\text{exp}(-M|t|)\psi$. Furthermore, since there is now only one zero-mode of $D^{1D}$, 
\begin{eqnarray}
\text{Ind}(\widehat{D}_{t})=\text{Ind}(D).
\end{eqnarray}
%さらに、Fig.1のようにシリンダーを曲げたもの (thick black line) と

The index \text{Ind}($D$) is the mod-two AS index on the semi-infinite bended cylindrical manifold which indicated as the black thick line in Fig.1. The previous work \cite{Fukaya:2019qlf} showed that this index is consistent with the mod-two AS index on the manifold $Y_{\text{cyl}}$.
Therefore, the formula (\ref{eq:indexcylinder}) tells us
\begin{eqnarray}
\label{eq:im}
\text{Ind}(\widehat{D}_{t})=\text{Ind}_{\text{APS}}(D|_{Y_{+}}).
\end{eqnarray}

\begin{figure}
\centering
\includegraphics[width=.5\textwidth]{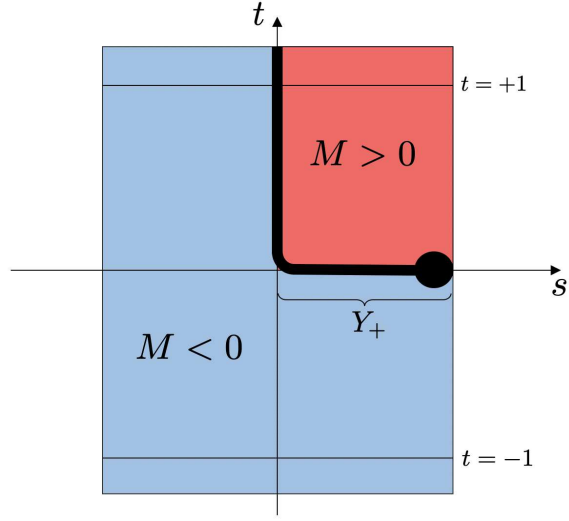}
\caption{Kink mass structure in $6$-dimensional plane}
\label{fig1}
\end{figure}

Next, let us rewrite the Dirac operator (\ref{eq:dirac6})
\begin{eqnarray}
&&\widehat{D}_{t}=\varepsilon(\partial_{t}+D_{t}),\\&&\varepsilon=\left(\begin{array}{cc}
1&0\\
0 &-1
\end{array}
\right),\ \ D_{t}=\left(\begin{array}{cc}
0&D+M\kappa(t)\\
-D+M\kappa(t) &0
\end{array}
\right).
\end{eqnarray}

 Let $\phi$ be the eigenfunction with the eigenvalue $\lambda_{t}$ of $D_{t}$. We assume that the $t$-dependent part of $\phi$ is $\exp[-\lambda_{t}t]$.  Then, the function $\phi$ is a zero-mode of $\widehat{D}_{t}$. From the normalization condition, the $\lambda_{t}$ must be negative when $t<0$ and positive when $t>0$. It means that there is a zero-mode of $\widehat{D}_{t}$ when the spectral flow occurs once:
\begin{eqnarray}
\label{eq:sf}
\text{Ind}(\widehat{D}_{t})=\text{Sf}(D_{-1},D_{1})\ \ (\text{mod} 2).
\end{eqnarray}

$D_{-1}$ and $D_{1}$ are evaluated as follows,
\begin{eqnarray}
D_{-1}=\left(\begin{array}{cc}
0&D_{PV}\\
D_{PV}^{\dagger} &0
\end{array}
\right),\ 
D_{1}=\left(\begin{array}{cc}
0&D_{DW}\\
D_{DW}^{\dagger} &0
\end{array}
\right).
\end{eqnarray}

For these operators, the operator $A$ which is given below satisfies $D_{1}=A^{\dagger}D_{-1}A$,
\begin{eqnarray}
A=\left(\begin{array}{cc}
0&D_{DW}\\
D_{PV}^{-1} &0
\end{array}
\right).
\end{eqnarray}

From (\ref{eq:spectral}), the spectral flow of $D_{t}$ is equivalent to the left hand side of (\ref{eq:modtwoDW}),
\begin{eqnarray}
\label{eq:ee}
\text{Sf}(D_{-1},D_{1})=\frac{1-\text{sgn\,Det}(D_{\text{DW}}D_{\text{PV}}^{-1})}{2}\ \ (\text{mod} 2).
\end{eqnarray}

The above formulae (\ref{eq:im}), (\ref{eq:sf}) and (\ref{eq:ee}) lead to our main result (\ref{eq:modtwoDW}).

\section{Separation of bulk and edge contributions of Mod-two APS index}
Finally,  we explain a natural decomposition of the mod-two APS index into the bulk and the edge contributions, and show the bulk/edge correspondence of the global anomaly. 

To this end, we introduce a single free domain-wall fermion with the opposite sign of mass, which does not contribute to the anomaly. The total partition function is given by
\begin{eqnarray}
\label{eq:anomalyinflow}
\text{Det}\left(\frac{D+M\epsilon(s)}{D-M}\right)\cdot\text{Det}\left(\frac{\partial -M\epsilon(s)}{\partial-M}\right)\propto (-1)^{I_{APS}^{mod-two}}.
\end{eqnarray}
In the previous section, it is proved that the phase of the product of these determinants is proportional to the mod-two APS index on the five-dimensional manifold. The left hand side of (\ref{eq:anomalyinflow}) is easily decomposed into the bulk and edge contributions as follows,
 \begin{eqnarray}
&&\text{Det}\left(\begin{array}{cc}
D+M\epsilon(s)&0\\
0&\partial-M\epsilon(s)
\end{array}\right) \cdot \text{Det}\left(\begin{array}{cc}
D-M&0\\
0&\partial-M
\end{array}\right)^{-1},\nonumber \\
&&=\text{Det}\left[\frac{\left(\begin{array}{cc}
D+M\epsilon(s)&0\\
0&\partial-M\epsilon(s)
\end{array}\right)}{\left(\begin{array}{cc}
D+M\epsilon(s)&\mu\\
\mu&\partial-M\epsilon(s)
\end{array}\right)}\right] \cdot \text{Det}\left[\frac{\left(\begin{array}{cc}
D+M\epsilon(s)&\mu\\
\mu&\partial-M\epsilon(s)
\end{array}\right)}{\left(\begin{array}{cc}
D-M&0\\
0&\partial-M
\end{array}\right)}\right], \nonumber \\
&&\equiv \text{Det}D^{\text{edge}}\cdot\text{Det}D^{\text{bulk}}.
\end{eqnarray}
In the second line, we insert the Pauli-Villars determinants in the denominator and numerator. In the hierarchical limit $\lambda_{\text{edge}}\ll \mu\ll M$ ($\lambda_{\text{edge}}$ is a typical energy scale of the edge modes), the first and second determinants are dominated by the edge-localized modes and the bulk modes, respectively. 

Since the Dirac operators $D^{\text{edge}}$ and $D^{\text{bulk}}$ are both real, the mod-two APS index is decomposed as follows:
\begin{eqnarray}
I_{APS}^{mod-two} &=&I_{\text{edge}}+I_{\text{bulk}}\ \  (\text{mod} 2),\\
I_{\text{edge}}&\equiv&\frac{1-\text{sgn}\,\text{Det} D^{\text{edge}}}{2},\\
I_{\text{bulk}}&\equiv& \frac{1-\text{sgn}\,\text{Det} D^{\text{bulk}}}{2}.
\end{eqnarray}

We remark that neither $\text{Det}D^{\text{edge}}$ nor $\text{Det}D^{\text{bulk}}$ are gauge-invariant due to the Pauli-Villars mass $\mu$. However the total partition function is gauge invariant. This is nothing but the bulk-edge correspondence for the global anomaly.

\section{Summary and Discussion}
%%じぶんのけんきゅうのまとめ
In this work, we have proved mathematically the equivalence between the formulation of the mod-two APS index with APS boundary condition and that with the domain-wall fermions. Our setup with the domain-wall fermions is equipped with many features consistent with the physics of topological materials such as the local boundary conditions.

Furthermore, we have found a natural separation of the bulk and edge contributions of the mod-two APS index. It shows a bulk-edge correspondence of the  global anomaly.

The application of our formulation to a lattice gauge theory is also interesting. In the $SU(2)$ gauge theory on a five-dimensional torus with fermions in the fundamental representations the lattice domain-wall Dirac operator is real,\begin{eqnarray}
D_{DW}^{lat}(x,y)=D_{W}(x,y)+\kappa m\delta_{x,y},
\end{eqnarray}
where $D_{W}$ is the Wilson-Dirac operator
\begin{eqnarray}
D_{W}=\sum_{\mu=1}^{5}\left(\gamma^{\mu}\frac{\nabla_{\mu}+\nabla_{\mu}^{\ast}}{2}-\frac{\nabla_{\mu}\nabla_{\mu}^{\ast}}{2}\right),
\end{eqnarray}
and $\nabla_{\mu}$ and $\nabla_{\mu}^{\ast}$ is forward/backward discrete derivatives, respectively. Then it is natural to define the lattice version of the mod-two APS index by
\begin{eqnarray}
\frac{1-\text{sgn\,Det}D^{lat}_{DW}}{2}\ \ (\text{mod} 2).
\end{eqnarray}
\\

We thank N. Kawai, Y. Kikukawa, Y. Kubota and K. Yonekura for useful discussions. This work was supported in part by JSPS KAKENHI (Grant numbers: JP15K05054, JP17H06461, JP17K14186, JP18H01216, JP18H04484, JP18K03620, 19J20559, JP20K14307, JP21K03222 and JP21K03574).

\end{document}